\documentclass[aps,twocolumn,showpacs,pra,amsmath,amssymb,floatfix,superscriptaddress]{revtex4-1}

\usepackage{color}
\usepackage{bbm}
\usepackage{graphicx}
\usepackage{dcolumn}
\usepackage{bm}
\usepackage{array}
\usepackage{float}
\usepackage{supertabular}
\usepackage{longtable}
\usepackage{mathrsfs}
\usepackage{txfonts}
\usepackage{wasysym}
\usepackage[T1]{fontenc}
\usepackage[usenames,dvipsnames]{xcolor}
\usepackage{amsmath}
\usepackage[colorlinks=true,citecolor=Cerulean,linkcolor=RubineRed,urlcolor=Cerulean]{hyperref}

\begin{document}

\title{Optimal preparation of the $W$ state for qubits with XY coupling}

\author{Dalton Jones}
\affiliation{Department of Physics and Astronomy, University of California, Los Angeles, California 90095, USA}
\affiliation{Department of Physics and Astronomy, Western Washington University, Bellingham, Washington 98225, USA}

\author{Armin Rahmani}
\affiliation{Department of Physics and Astronomy, Western Washington University, Bellingham, Washington 98225, USA}
\affiliation{Advanced Materials Science and Engineering Center, Western Washington University, Bellingham, Washington 98225, USA}
\affiliation{Kavli Institute for Theoretical Physics, University of California, Santa Barbara, California
93106, USA}

\date{\today}

\begin{abstract}

Using simulated annealing, we find optimal protocols that evolve a simple product state into a three-qubit $W$ state with a Hamiltonian that describes XY coupling and single-qubit gates, and determine the associated quantum speed limit. Applying Pontryagin’s minimum principle, we fully characterize the optimal “bang-bang” protocols. While leakage affects performance, the protocols remain robust to implementation errors and operate well within relaxation and decoherence times. Our findings highlight Pontryagin’s principle as a powerful tool for designing pulse shapes that directly link device interactions to specific quantum gates and target states.

\end{abstract}

\maketitle

\section{Introduction}

Numerous advances have been made in developing coherent quantum systems~\cite{coh2, coh3, coh4} for information processing. Since entanglement is a key resource in quantum information processing~\cite{nielsen,ladd}, it is desirable to create entangled states as quickly as possible. In three-qubit systems, there are two inequivalent classes of entanglement: the Greenberger-Horne-Zeilinger (GHZ) states, which exhibit maximal multipartite entanglement, and the $W$ states, which possess maximal bipartite entanglement among all qubit pairs~\cite{Tamaryan,dur} and are notably robust to noise and particle loss~\cite{dur,WlikeEnt}.

In this work, we focus on the optimal preparation of the standard $W$ state at the quantum speed limit, although similar methods apply to GHZ and other $W$-type entangled states. The creation of entangled states has been explored in various platforms~\cite{chenY,anN,zhangS,kangY,eiblM,zohX,Watts,Goerz,ManyBodyW,TrappedIon,PhotGHZ}, including the realization of three-qubit $W$ states of superconducting qubits~\cite{3Q} and photonic qubits~\cite{PhotW}.
However, the problem of determining the fastest possible unitary operation that creates a specific $W$ state under a fixed Hamiltonian remains open. Here, we address this challenge using optimal control theory and Pontryagin’s minimum principle.

For our study, we consider a three-qubit Hamiltonian with generic tunable single-qubit fields and XY couplings. Note that while experimentally relevant parameter estimates are made with respect to the gmon superconducting qubit architecture~\cite{Mart}, our results are relevant to quantum dot electron-spin qubits~\cite{ElecDot,ElecDotiSWAP}, trapped ion qubits~\cite{TrapIon}, and others with similar Hamiltonians. This structure enables precise, time-dependent control of interacting qubits.
A common approach to generating maximally entangled states is adiabatic evolution, where the desired state is reached over timescales much longer than the inverse energy gap between the ground and first-excited states (throughout this work, references to these states pertain to a fixed symmetry sector of the system’s Hamiltonian). However, if the chosen time-dependent Hamiltonian trajectory includes a level crossing—or an extremely small gap—between these states, adiabatic evolution fails. In cases of exact level crossings, the initial and final states may belong to different symmetry sectors, preventing evolution from one to another with a symmetry-preserving Hamiltonian. To overcome this limitation, we employ optimal control.

 Optimal control techniques that shortcut the adiabatic evolution have been studied extensively~\cite{OCreview, OCrev2, OC1, OC3, OC4, OC6, OC7, OC8, OC13, OC14, OC16, manybody1, manybody2, manybody3, fewbody1, fewbody2, top3, top5, OCcold1, OCcold2,cool,fewbody1, fewbody2,OCqs}.
Previously, we found optimal protocols for evolving a two-qubit system from a product state to a singlet state~\cite{OP}. These solutions, confirmed by Pontryagin's minimum principle, were bang-bang, i.e., a sequence of square pulses. They substantially outperform the adiabatic evolution by converging at the quantum speed limit of the system. Here, we generalize the results of Ref.~\cite{OP} to a system of three qubits. We find optimal solutions that prepare an entangled state, known as the $W$ state,
\begin{equation}\label{eq:W}
|W\rangle=\frac{1}{\sqrt{3}}(|\uparrow\uparrow\downarrow\rangle + |\uparrow\downarrow\uparrow\rangle + |\downarrow\uparrow\uparrow\rangle),
\end{equation}
from a product state over a range of allowed times. We first perform numerical optimizations to minimize a cost function defined by the overlap between the system’s final state and the target entangled state \eqref{eq:W}. Pontryagin’s minimum principle is then applied to verify and refine these solutions. Finally, we examine the effects of implementation errors and leakage.

The protocols show strong robustness against pulse height and timing errors. However, leakage in systems such as superconducting qubits, that have a small anharmonicity relative to single-qubit transition frequencies~\cite{1Qleak,gmonrelax,Mart2}, presents a significant challenge at the timescale at which our protocols operate.

Our results provide explicit protocols for optimally controlling three qubits to prepare the $W$ state \eqref{eq:W} from an initial product state with the same symmetry in the shortest possible time. These findings are consistent with the quantum speed limit that bounds the minimal time to reach a target state~\cite{QSL}. They also advance our understanding of the structure of optimal protocols, extending insights from earlier work on two-qubit systems~\cite{OP}. We find that the optimal solutions in both cases share key features: they both exhibit solutions with bang-bang structure, characterized by a small number of switching points. Additionally, both systems display singular intervals that, in principle, permit deviations from the bang-bang form. However, in both cases, the optimal protocols ultimately converge to a bang-bang structure. It appears that the complexity of optimal protocols that prepare entangled states scales with the number of qubits in the system.

\section{Model and setup}

Superconducting~\cite{Mart}, quantum dot~\cite{ElecDot,ElecDotiSWAP}, trapped ion~\cite{TrapIon}, and other qubit architectures support tunable single-qubit fields $(B_{i,x},B_{i,y},B_{i,z})$ and (effective) $XY$ qubit couplings. For three qubits, the Hamiltonian has twelve tunable parameters, including 9 single-qubit fields and qubit-qubit couplings $J_{12},J_{23},J_{31}$:
\begin{equation}\label{eq:hamil}
H = H_B-H_J = \sum_{i=1}^3 \mathbf{B}_i\cdot\mathbf{\sigma}_i - \sum_{\langle m,n\rangle}J_{mn}(\sigma^x_m \sigma^x_{n} + \sigma^y_m \sigma^y_{n}).
\end{equation}
The index $i$ represents the $i$th qubit, the sum over $\langle m,n\rangle$ includes the three nearest neighbor pairings in the triangular formation, and $\mathbf{\sigma}_i$ is the Pauli vector. Thus, we allow all parameters to be tuned as a function of time in the range:
\begin{equation}\label{eq:range}
0\leqslant  B_{i,x}(t),B_{i,y}(t),B_{i,z}(t),J_{12}(t),J_{23}(t),J_{31}(t) \leqslant \Lambda,
\end{equation}
where $\Lambda$ sets the energy scale of the parameters. In practice, the energy scale for couplings is smaller than that of the single qubit fields. However, they typically do not differ by several orders of magnitude~\cite{Mart} and, values of $B/2\pi = J/2\pi = 50\ {\rm MHz}$ were used in Ref.~\cite{Spect}. We use natural units ($\hbar=1$) and set $\Lambda=1$.

Our target state is $|\psi_{\rm target}\rangle = |W\rangle$ of Eq.~\eqref{eq:W}, where the spin-up and spin-down states are eigenstates of $\sigma^z_i$ with eigenvalues $m_i = \pm 1$, respectively. The above state corresponds to the ''standard'' $W$ state commonly used in the literature and is adopted here for concreteness. We could construct other states exhibiting the same $W$-type entanglement by introducing arbitrary phase factors in the amplitudes of the basis states, and apply a similar method to prepare them.

{\bf \textit {Adiabatic evolution.}} If we set the $x$ and $y$ components of the $B$ field to zero, the Hamiltonian \eqref{eq:hamil} preserves the total spin in the $z$ direction. The target state has $\sum_{i=1}^3m_i=+1$. In the three-dimensional Hilbert space corresponding to this sector, the Hamiltonian can be written as
\begin{equation}\label{eq:3ham}
H=
\left(\begin{array}{cccc}
    B_1+B_2-B_3  & -2J_{23} & -2J_{31} \\
    -2J_{23}  & B_1-B_2+B_3 & -2J_{12} \\
    -2J_{31}  & -2J_{12} & B_2-B_1+B_3
\end{array} \right),
\end{equation}
where we have omitted the $z$ subscript on the $B_{j,z}$ parameters. The target state is a ground state of the Hamiltonian \eqref{eq:hamil} for, e.g., $J_{12},J_{23},J_{31}=1$ and $B_1,B_2,B_3=0$. For $J_{12},J_{23},J_{31}=0$ and $B_1,B_2,B_3=1$, the three qubits are decoupled, and the degenerate ground states (in the sector with $\sum_i m_i=+1$) are 
\begin{equation}
|\psi_1\rangle = |\uparrow\uparrow\downarrow\rangle,\quad |\psi_2\rangle=|\uparrow\downarrow\uparrow\rangle,\quad |\psi_3\rangle=|\downarrow\uparrow\uparrow\rangle.
\end{equation}
These unentangled direct-product states are easy to prepare. We choose one of these ground states, namely $|\psi_1\rangle$, as our initial state $|\psi(0)\rangle = |\psi_1\rangle$.

\begin{figure}
  \par\medskip
  \includegraphics[width=\linewidth]{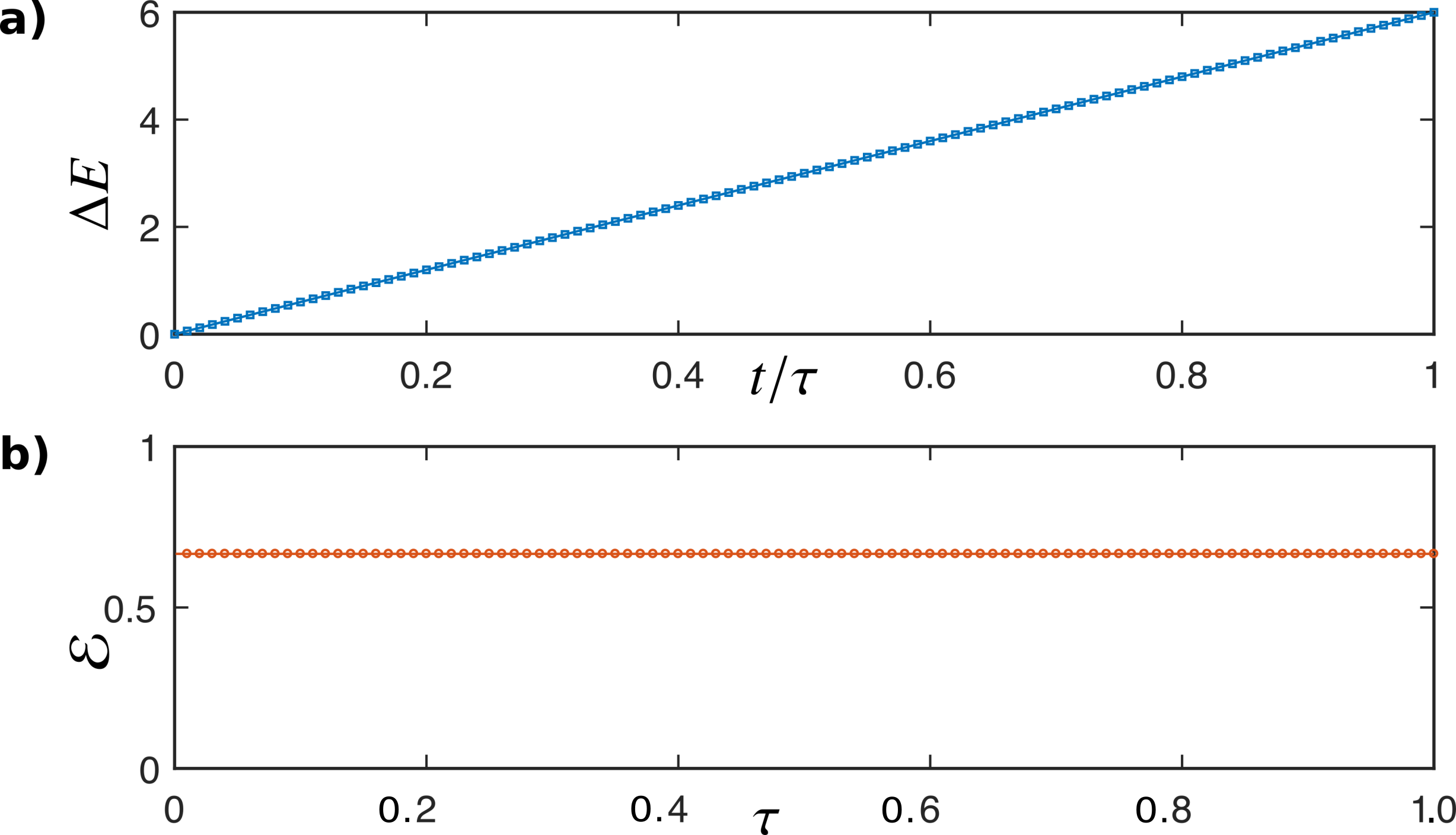}
  \caption{\textbf{a)} The energy gap for $B_1=B_2=B_3=1-t/\tau$ and $J_{12}=J_{23}=J_{31}=t/\tau$ as a function of $t/\tau$ indicating the initial degeneracy. \textbf{b)} The error function ${\cal E}=2/3$ for the final state is independent of the total time, showing the failure of the adiabatic transformation.}
  \label{fig:Evalues}
\end{figure}

As an example, we consider an adiabatic trajectory that evolves all the $B$ parameters according to $1-t/\tau$ and all the $J$ parameters according to $t/\tau$ from $t=0$ to $t=\tau$. The energy eigenvalues are $1-5t/\tau$ and doubly degenerate $1+t/\tau$. Therefore, the energy gap $\Delta E= 6t/\tau$ between the ground states and the first excited state along this trajectory (in the 3-dimensional sector), shown in Fig.~\ref{fig:Evalues}(a), vanishes at $t=0$. We quantify the difference between the final state of the system and the target state by the cost function
\begin{equation}\label{eq:cost}
{\cal E}=1- |\langle \psi_{\rm target}|\psi(\tau)\rangle|^2.
\end{equation}
This error is nonnegative, vanishes if we reach the target state, and has an initial value $\mathcal{E}(\tau =0) = 1 - |\langle W|\psi(0)\rangle|^2 = 2/3$.

In Fig.~\ref{fig:Evalues}(b), we plot the error as a function of the total time $\tau$. The adiabatic evolution does not change the error $\cal E$ from its initial value of $2/3$. The Hamiltonian with an initial state $|\psi_1\rangle$, leads to normalized final states of the form $|\psi(\tau)\rangle=(\alpha-i\beta, 2i\beta, 2i\beta)$, where $\alpha$ and $\beta$ are real numbers. All of these final states have the same error $\cal E$ with respect to the $W$ state. It is not surprising that we cannot reach the target state adiabatically because the initial state is one of three degenerate ground states, and the initial energy gap is zero.

{\bf \textit {Intermediate evolution to $W$-like state.}}  A $W$-like state can be achieved by evolving the initial state $|\psi(0)\rangle = |\psi_1\rangle$, according to the Hamiltonian with all parameters set to zero except $J_{23} = J_{31} = 1$, over a period of time $t_0 = \frac{1}{2\sqrt{2}}\cos^{-1}(\frac{1}{\sqrt{3}}) \approx 0.34$. The $W$-like state is $|\psi(t_0)\rangle = (1,i,i)/\sqrt{3}$ which may be rotated to the desired $W$-state via a $\frac{\pi}{2}$ phase rotation of the $|\psi_2\rangle$ and $|\psi_3\rangle$ basis states. This is achieved by setting all $J$ parameters to zero, $B_1 = B_2 = \frac{1}{2}$, and $B_3 = 1$ and evolving the state by $t=\frac{\pi}{2} \approx 1.57$. The total time to reach the specific target $W$ state in this case is $t\approx 1.9$. This protocol provides a useful result for comparison, and highlights the dependence of the minimal preparation time on the choice of the $W$-type state. Our goal is to reach the specific $W$-state in the shortest possible time. This study can be repeated with minimal change for other $W$-type states.

\section{Numerical Optimization}\label{sec:numOpt}

To find the shortest time required to reach the $W$ state~\eqref{eq:W}, we scan over a range of final times $\tau$ (starting from zero) and determine the minimal error \eqref{eq:cost} between the initial and final states at each final time until the error converges to zero at $\tau_C$. The error is a functional of the twelve time-dependent control parameters $B_{n,k}(t)$ and $J_{mn}(t)$ in the 8-dimensional Hamiltonian \eqref{eq:hamil}, which can be tuned to arbitrary functions of time as long as they remain in the range specified in Eq.~\eqref{eq:range}. We simultaneously minimize the error with respect to the trajectories of all twelve parameters, assuming no symmetry.

{\bf \textit {Piecewise-constant optimization.}} In a brute-force approach, the error functional must be converted into a multivariable function. Following the strategy of the quantum approximate optimization algorithm (QAOA)~\cite{QAOA}, we divide the total evolution time $\tau$ into $N$ equal segments of duration $\tau/N$. In the limit $N \to \infty$, any admissible control parameter can be represented as a piecewise-constant function over these intervals. By starting with a finite $N$ and gradually increasing it, we can infer the overall structure of the optimal protocol and subsequently obtain numerically exact results through a secondary optimization based on Pontryagin’s minimum principle. We then write the final state as
${|\psi(\tau)\rangle}=U_N(\tau-t_{N-1})U_{N-1}(t_{N-1}-t_{N-2})....U_1(t_1){|\psi(0)\rangle}$,
where $U(t)$ is the unitary time evolution operator. During each of these intervals, the $B_{n,k}(t)$ ($n=1, 2, 3$ and $k = x,y,z$) and $J_{mn}(t)$ ($mn=12, 23, 31$) parameters are allowed to be any constant value within the specified range~\eqref{eq:range}. The value of these parameters for all intervals determines the unitary evolution matrices, which in turn determine the error $\cal E$ of the protocol as a function of $12N$ controls.

\begin{figure*}
  \par\medskip
  \includegraphics[width=0.86\linewidth]{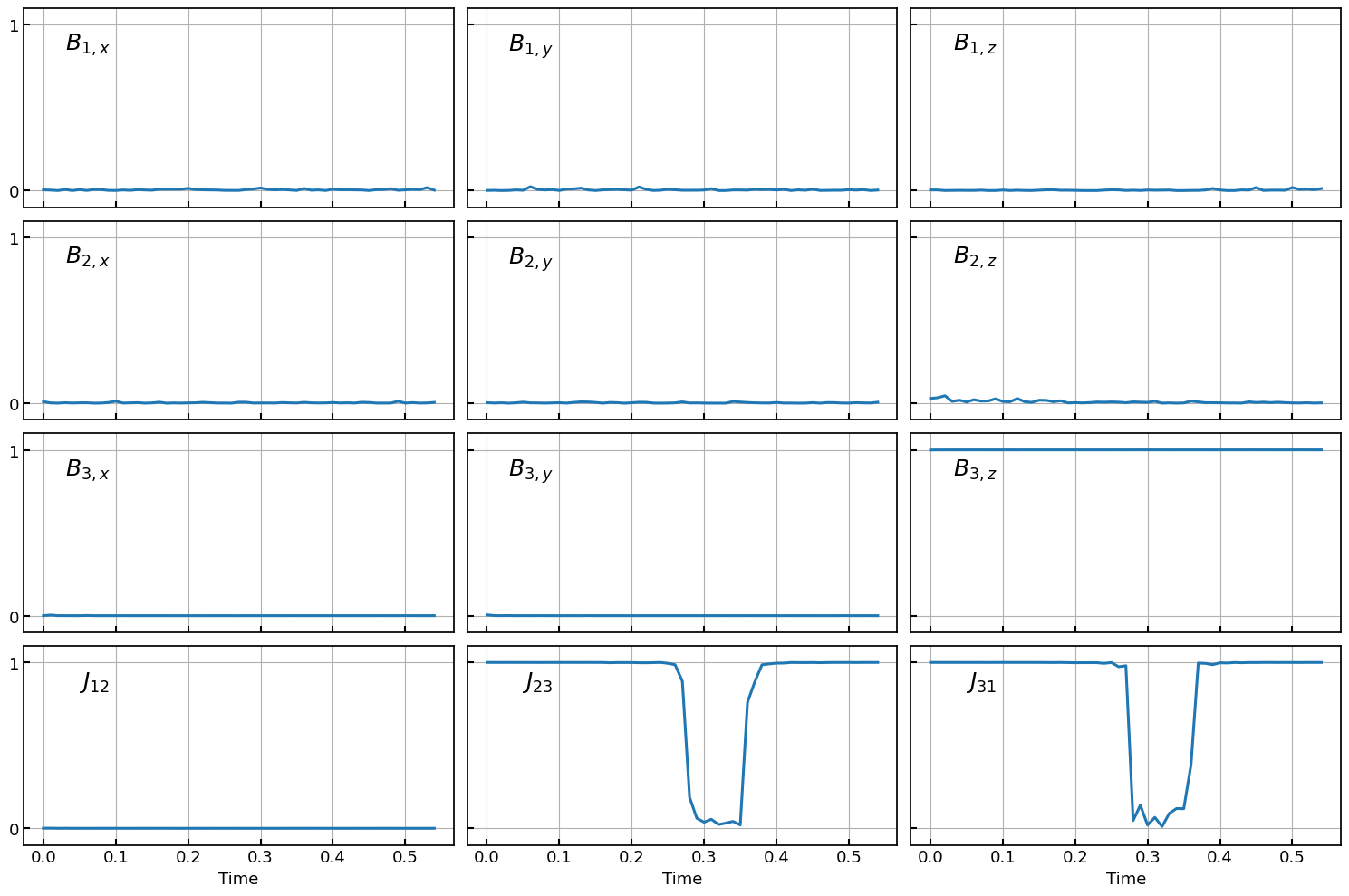}
  \caption{Protocols for the $B$ and $J$ parameters determined by the $N$-interval optimization with $N=55$, a final time $\tau=0.55$, and an error of ${\cal E} \approx 0.0492$.}
  \label{fig:nstep}
\end{figure*}

We then use a simulated annealing algorithm (a global optimization algorithm designed to escape local minima) to minimize the error with respect to the $12N$ parameters. As expected, these optimizations converge to bang-bang protocols with minor distortions due to coarse discretization. For example, the protocol shown in Fig.~\ref{fig:nstep} has an optimized error of ${\cal E} \approx 0.0492$ for a final time $\tau=0.55$. The shapes of the protocols clearly indicate that $B_{3,z} = 1$ is the only non-zero $B$ parameter, $J_{12} = 0$, and that $J_{23} = J_{31}$ is the only non-trivial protocol with two discontinuous jumps. The jumps in the $J_{12} = J_{23}$ protocol appear to happen at $t_1 \approx 0.29$ (from 1 to 0) and $t_2 \approx 0.36$ (from 0 to 1). We then performed bang-bang optimization by allowing a maximum of three jumps to account for a potential third jump that may have been missed.

{\bf \textit {Bang-Bang optimization.}} The bang-bang optimization is designed to refine these solutions by allowing the protocols to undergo switches, at optimized (non-discrete) values of time, between their maximum or minimum values (1 or 0). Based on the results of the optimization of the $N$-step, which did not find more than two switches, we allow a maximum of three switches (and indeed find only two upon convergence). The $x$ and $y$ components of the single qubit fields are also set to zero based on the results of the $N$-step optimization.
As these fields break the $\sum_i m_i = +1$ symmetry of the initial and final states, it is not surprising that they are not turned on in the optimal protocols.
We use the notation where $B_1,B_2,B_3$ represent the $z$ components of each single qubit fields.

\begin{figure}
  \par\medskip
  \includegraphics[width=0.9\linewidth]{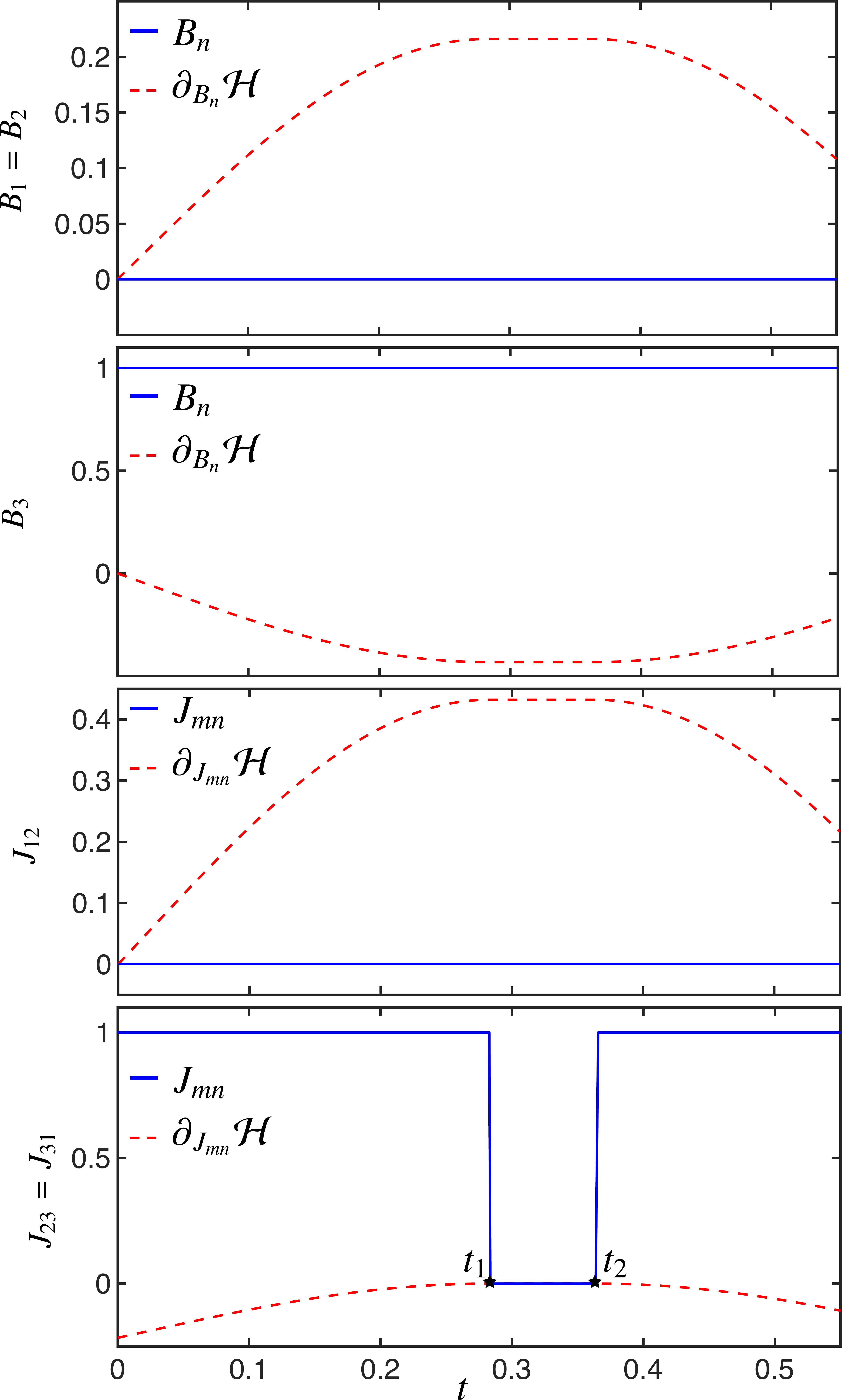}
  \caption{Optimal protocols (from the bang-bang optimization) for all control parameters with final time $\tau=0.55$, error ${\cal E} \approx 0.0468$, $t_1 \approx 0.283$, and $t_2 \approx 0.369$. The dashed lines represent the OCH coefficients.}
  \label{fig:Protocol}
\end{figure}

We again use a simulated annealing algorithm to minimize the error function $\cal E$. The optimal protocols we obtained corresponded to a slightly lower error than those found by the $N$-step optimization, which did not allow for the fine-tuning of switching times. An example of a solution for $\tau=0.55$ is shown in Fig.~\ref{fig:Protocol}. The solid blue lines indicate the Hamiltonian parameter as a function of time. In this particular case, the parameters $B_1$, $B_2$, and $J_{12}$ are confirmed to be turned off, $B_3$ is on, and the protocol for $J_{23}=J_{31}$ is also confirmed to have two switches at $t_1 \approx 0.283$ and $t_2 \approx 0.369$ during evolution. The error optimized with respect to the bang-bang protocol at final time $\tau=0.55$ is ${\cal E} \approx 0.04908$, which shows a slight improvement (of order $10^{-4}$) over the optimal error found by the $N$-step optimization (${\cal E} \approx 0.0492$), Fig.~\ref{fig:nstep}.

The solutions we found over a range of final times ($0.1\leq \tau \leq 0.77$) are split into two groups based on the differences between their $J_{23} = J_{31}$ protocols. For $\tau > \tau_0 \approx 0.469$, the optimal solutions have the same general structure as our example at $\tau=0.55$ (Fig.~\ref{fig:Protocol}). For $\tau \leq \tau_0 \approx 0.469$, the switching times converge $t_1=t_2$, such that the resulting protocol is constant $J_{23} = J_{31} = 1$ during evolution. The minimum error over time is plotted in Fig.~\ref{fig:error_switches}(a) and depicts convergence to zero as $\tau \rightarrow  \tau_c \approx 0.77$, where $\mathcal{E} \approx 10^{-5}$. The switching times that define the optimal protocols are plotted in Fig.~\ref{fig:error_switches}(b) and show that $t_1 \approx 0.283$ is independent of the final time $\tau$ while $t_2 = t_1$ until $\tau_0 \approx 0.47$, where $t_2$ begins increasing linearly with $\tau$. 

\section{Pontryagin's Minimum Principle and verification of the optimal protocols}

To apply the formalism of Pontryagin's Minimum Principle (see the Supplementary Material for a review of the formalism) to the problem at hand, we start with the optimal control Hamiltonian (OCH) 
\begin{equation}{ \cal H}= {\rm Im}[\langle\Pi(t)| H(t) |\psi(t)\rangle]\label{eq:OCH2}
\end{equation}
 written in terms of the conjugate momenta $|\Pi(t)\rangle$, defined as variables conjugate to the dynamical variable encoded in the state $|\psi(t)\rangle$. These momenta are in fact Lagrange variables imposing the Schr\"odinger equation as a constraint.

 The equations of motion for the conjugate momenta (see Eq. (3) of the Supplementary Material)  $|\Pi\rangle$ are also the Schrodinger equation 
$\partial_t|\Pi(t)\rangle=-iH(t)|\Pi(t)\rangle$.
Unlike the quantum state $|\psi(t)\rangle$, for which the boundary conditions are given at the initial time $t=0$, Pontryagin's theory specifies a boundary condition for $|\Pi(t)\rangle$ at the final time $t=\tau$, which depends on the cost function.
Using the error function~\eqref{eq:cost} with $\langle \psi_{\rm target}|$ represented by $(1/
\sqrt{3},1/\sqrt{3},1/\sqrt{3})$, we write the cost function as
\begin{equation}
{\cal E}=1-{1\over 3}\langle \psi(\tau)|{\cal M}|\psi(\tau)\rangle, \quad {\cal M}\equiv
\left(\begin{array}{cccc}
1 & 1 & 1 \\ 
1 & 1 & 1 \\ 
1 & 1 & 1
\end{array} \right).
\end{equation}
The above expression leads to the boundary condition
\begin{equation}\label{eq:piBC}
|\Pi(\tau)\rangle=-{2 \over 3}{\cal M}|\psi(\tau)\rangle.
\end{equation}
The above expression follows from the fact that the final values of the conjugate momenta are equal to the derivatives of the cost function with respect to the corresponding dynamical variable.

If we have a dynamical protocol for the Hamiltonian parameters $B_n(t)$ and $J_{mn}(t)$, we can determine $|\psi(t)\rangle$ for all times from the known initial conditions for $|\psi\rangle$. Then the boundary condition \eqref{eq:piBC} gives the final value of $|\Pi\rangle$, and we can solve the Schr\"odinger equation backward in time and find $|\Pi(t)\rangle$. Then, using Eq.~\eqref{eq:OCH2}, we obtain the functional dependence of the OCH on the Hamiltonian parameters.

Importantly, since the elements of $H(t)$ have a linear dependence on $B_n(t)$ and $J_{mn}(t)$, the OCH~\eqref{eq:OCH2} is a linear function of these parameters. The coefficient of the control parameter $f$ in OCH is given by $\partial_f \mathcal{H}$. Pontryagin's theory states that the optimal $f$ must minimize the OHC at all times. Thus, we generically expect a bang-bang protocol for the control parameter, as it takes its minimum or maximum alowed value depending on the sign of the coefficient above. For example, the coefficient of $B_1$ is
\begin{equation}\label{eq:coeff}
\partial_{B_1} {\cal H}={\rm Im}[\langle\Pi(t)| {\cal K}_{B_{1}} |\psi(t)\rangle],
\end{equation}
where ${\cal K}_{B_1}$ denotes the partial derivative of the Hamiltonian~\eqref{eq:3ham} with respect to $B_1$. In this particular case, it is a diagonal matrix with elements +1, -1 , and -1 on the diagonal. If this coefficient is positive (negative), we must set $B_1$ to its minimum (maximum) allowed value $B_1=0$ ($B_1=\Lambda$). The only exception to this generic rule is if the coefficient has a finite singular interval over which the coefficient identically vanishes. The OCH coefficients and optimal values of all other parameters can be constructed according to the same process outlined above \eqref{eq:coeff}.

Pontryagin's theorem provides a powerful iterative approach to finding the numerically exact optimal protocol. To confirm and further refine the protocols found using the simulated annealing algorithm, we used the Pontryagin's minimum principle.  
As stated before, we can check whether we have found the optimal protocol by solving for the OCH, $\cal H$, and  the coefficients such as $\partial_{J_{23}}{\cal H}={\rm Im}[\langle\Pi(t)| {\cal K}_{J_{23}} |\psi(t)\rangle]$ \eqref{eq:coeff} as a function of time for a given protocol. To minimize the OCH, we must keep the parameter at its maximum (minimum) value while its coefficient is negative (positive). Thus, the coefficients for each $B_n$ and $J_{mn}$ parameter in the OCH should pass through zero at the same time that the parameter switches between maximum and minimum values.

In Fig.~\ref{fig:Protocol}, we show the corresponding coefficient (dashed red line) to each of the control parameters, calculated with the method above (see the Supplementary Material for more details). All results are consistent with the Pontryagin's theorem. For the constant protocols $J_{12}$ and $B_{1,2,3}$, the sign of the coefficient agrees with the value of the control parameter. For $J_{23}=J_{31}$, the coefficient is negative when the control parameter is equal to one. Interestingly, the finite interval with a vanishing control parameter has a coefficient that is identically zero. Similar behavior occurs for the optimal preparation of the maximally entangled state of two qubits; there are \textit{singular} intervals, where the Pontryagin's theorem does not determine the protocol. A priori, optimal protocols do not need to be bang-bang if singular intervals are present. However, in this case, the optimal protocol is nevertheless bang-bang.

We found that for $0\leqslant\tau\leqslant\tau_0=0.469017$, all the control parameters are constant through the entire evolution. For larger $\tau$, a single dip appears in $J_{23}$ and $J_{31}$ at time $t_1=0.282687$. A second jump appears at a time $t_2$ with $t_1\leqslant t_2<\tau \ $. We can put these two patterns on the same footing by setting $t_2=t_1$ for $0\leqslant\tau\leqslant\tau_0$. All intervals are nonsingular except for the $t_1<t<t_2$ singular interval. 
\begin{figure}
  \par\medskip
  \includegraphics[width=0.9\linewidth]{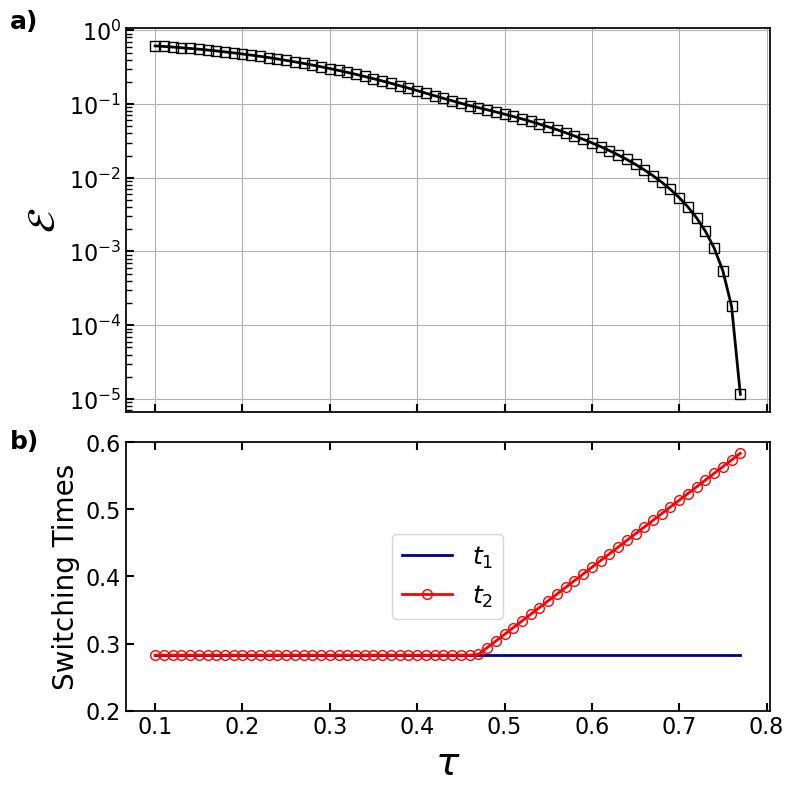}
  \caption{The optimal protocols are characterized over a range of final times ($0.1 \leq \tau \leq 0.77$) by \textbf{a)} the minimum error $\mathcal{E}$ between the final state and the entangled $W$ state and \textbf{b)} the switching times $t_1$ and $t_2$ that define the optimal bang-bang protocol of $J_{23} = J_{31}$ parameters.}
  \label{fig:error_switches}
\end{figure}
While $t_1$ is independent of $\tau$, as seen in Fig.~\ref{fig:error_switches}(b), $t_2$, the time of the second switch in the $J_{23} = J_{31}$ protocol, depends linearly on $\tau$. We quantified this relationship by investigating the evolution of the OCH coefficients of these parameters. The switching time $t_2$ as a function of $\tau$ is 
\begin{equation}\label{eq:rel}
t_2(\tau)=(\tau-\tau_0)+t_1 \quad (\tau\geqslant\tau_0),
\end{equation}
where we found found a slope of 1 by performing a linear fit on the numerical data shown in Fig.~\ref{fig:error_switches}b.

The pattern shown in Fig.~\ref{fig:Protocol} together with Eq.~\eqref{eq:rel} and the fixed values of $t_1$ and $\tau_0$ uniquely determine the optimal protocol for any total time $\tau$. The next question is, what is the minimum $\tau$, for which the optimal protocol can prepare the $W$ state exactly. By scanning over a range of final times, we have determined that the desired maximally entangled state is reached in a finite amount of time. As seen in Fig.~\ref{fig:error_switches}, for $\tau\geq\tau_c\approx 0.77$, the error between the final state and the desired state becomes negligible (of order $10^{-5}$). Since we know the bang-bang pattern of the optimal protocol, we can determine the precise time $\tau_c$ at which the error vanishes by writing the error $\cal E$ as an analytic function of $\tau$. The calculation, shown in the Supplementary Material, leads to $\tau_c=0.7727$. Thus, the shortest amount of time required for the bang-bang protocol to exactly reach the entangled W state of three qubits is of order unity in natural units, given a maximum energy $\Lambda = 1$~\eqref{eq:range}. The sudden convergence of the system to the desired state in Fig.~\ref{fig:error_switches} at time $\tau_C$ is consistent with the presence of a quantum speed limit~\cite{QSL}.

\section{Robustness of Optimal Solutions}

\begin{figure}
  \par\medskip
  \includegraphics[width=0.9\linewidth]{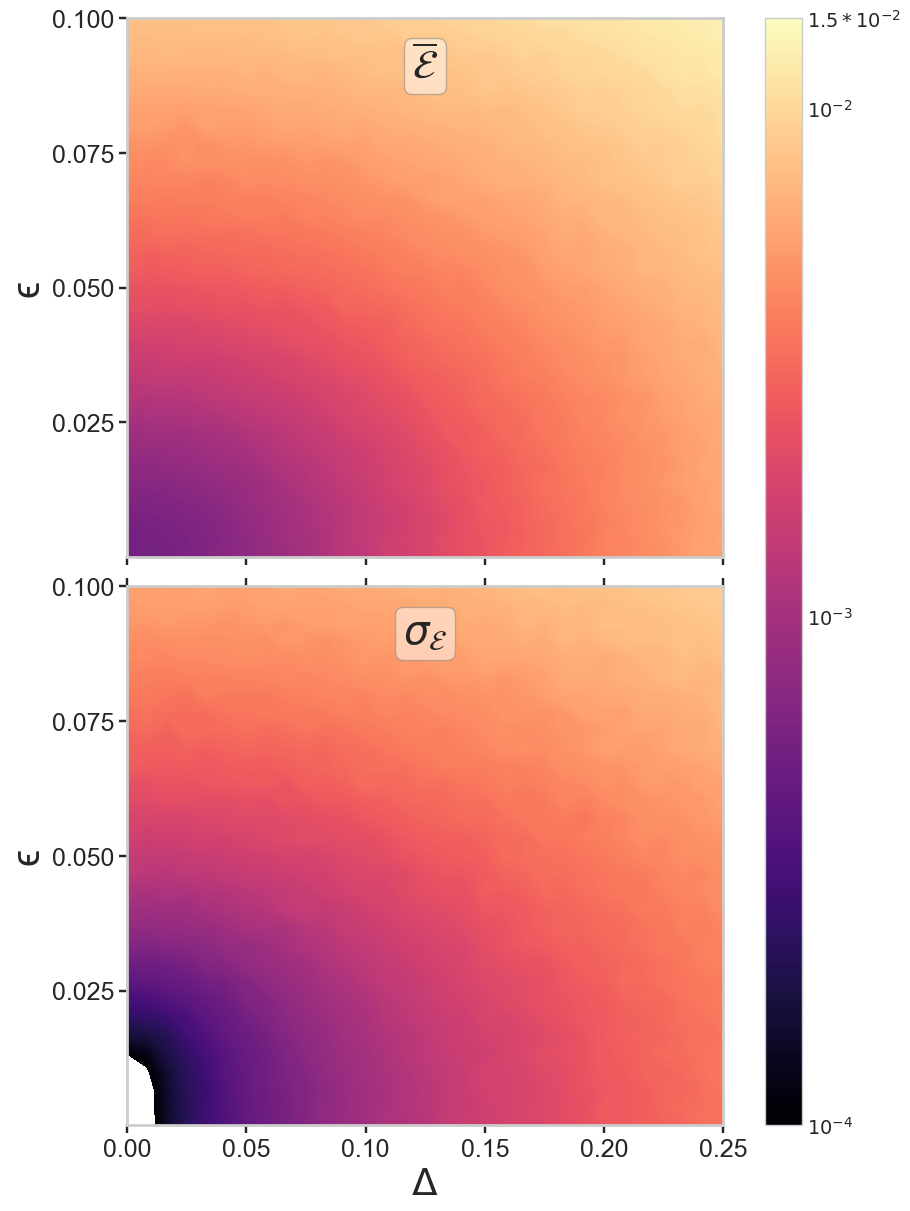}
  \caption{The average final error $\overline{\mathcal{E}}$ and its standard deviation $\sigma_\mathcal{E}$ resulting from $10^3$ realizations of randomly sampled errors from a uniform distribution, defined by the range $[-\epsilon/2,\epsilon/2]$ for timing errors and $[-\Delta/2,\Delta/2]$ for pulse height errors. Note that values below $10^{-4}$ are outside of the log-scale color range, such that the plot is unable to assign a color to the values of $\sigma_\mathcal{E}$ near $\Delta = \epsilon = 0$.}
  \label{fig:stability}
\end{figure}

When implementing these optimal protocols in an experimental setting, there will be some unavoidable sources of error, namely timing inaccuracies, pulse height inaccuracies, relaxation, decoherence, and leakage. Thus, it is essential to investigate these effects on the final error of the optimal protocol. Here we focus on the first two sources, leaving the other sources to the Supplementary Material.

We introduced random switching time and pulse height errors drawn from a uniform distribution ranging from $[-\epsilon/2,\epsilon/2]$ and $[-\Delta/2,\Delta/2]$, respectively. The randomized errors were applied independently to each switching time and non-zero pulse height (meaning independent errors are applied to $B_{3}$ and each non-zero interval in the protocols for $J_{23}$ and $J_{31}$). We work with the optimal protocols found at $\tau=0.75$, where the final error for a perfect protocol is ${\cal E} \approx 5.47\times10^{-4}$.  Upon perturbing the optimal protocol by these implementation errors, we calculate the final error, $\cal E$, for each realization. We found that $10^3$ realizations are enough to obtain convergence for both the average final error $\overline{\cal E}$, and the standard deviation $\sigma_\mathcal{E} \ $, of the final error. Here, the overline indicates the average over all realizations of implementation inaccuracies, and  $\sigma=\sqrt {\overline{{\cal E}^2}-\left(\overline{\cal E}\right)^2}$.

The results of these calculations are shown in Fig~\ref{fig:stability}. Both the average error and its standard deviation grow slowly in $\epsilon$ and $\Delta$. The average error increases by $10\%$ of the minimum error ($\overline{\mathcal{E}} = 1.1\mathcal{E}$) for $\Delta \approx 0.027$ with a standard deviation of $\sigma_\mathcal{E}(\Delta, \epsilon=0) \approx 0.4\mathcal{E}$ or $\epsilon \approx 0.01$ with a standard deviation of $\sigma_\mathcal{E}(\Delta=0,\epsilon) \approx 0.01\mathcal{E} \ $. Combining the timing and pulse height errors gives approximately a $20\%$ increase in the average error with respect to the minimum error ($\overline{\mathcal{E}} \approx 1.2\mathcal{E}$) and a standard deviation of $\sigma_\mathcal{E}(\Delta,\epsilon) \approx 0.4\mathcal{E} \ $. Thus, we have shown that for error realizations, where $\Delta$ is $2.7\%$ of the system's energy scale $\Lambda$ and $\epsilon$ is $3.5\%$ of the first switching time $t_1$, the system yields negligible errors of the order $10^{-4}$ in the final wavefunction. This leaves plenty of room for experimental error when implementing these optimal protocols. As discussed in the Supplementary Material, the robustness to $\epsilon$ allows for creating ramps of width $\epsilon$ instead of sudden jumps to exclude high-frequency modes that cause leakage.

\section{Conclusion}

We started with a tunable Hamiltonian~\eqref{eq:hamil}, with generic single-qubit fields and $XY$ qubit couplings, that describes certain regimes within superconducting gmon~\cite{Mart}, quantum dot electron-spin~\cite{ElecDot,ElecDotiSWAP}, trapped ion~\cite{TrapIon}, and other qubit architectures. Using this Hamiltonian structure, we identified optimal control protocols that prepare the three-qubit $W$ state from an easily prepared product state within a finite time. These protocols take the form of simple bang-bang sequences, in which control parameters are switched on and off at precisely optimized times. Each “bang” corresponds to a distinct unitary operation which has a simple and straight forward experimental implementation. The total convergence time of our protocols can be reduced by increasing the system’s energy scale~\eqref{eq:range}, and order-of-magnitude estimates show that, for example, these protocols operate on timescales far shorter than typical relaxation and decoherence times for gmon qubits.

The protocols are also notably robust to realistic experimental noise and parameter fluctuations. This robustness allows their implementation alongside other operations before decoherence sets in, making them highly suitable for high-fidelity quantum information processing. However, for systems that suffer from leakage, the structure of optimal protocols may be modified and the convergence time may be extended. Comparing our findings with those of Bao et al.~\cite{OP}, we find that the optimal three-qubit protocols converging to the $W$ state include one additional switching event compared to those that maximally entangle two qubits. In both two- and three-qubit systems, the bang-bang character of the optimal control persists even when the OCH coefficient becomes singular~\cite{OP}, although such singularities may complicate proofs of optimality for larger systems.

Future work will extend these results to include leakage effects, alternative initial states, systems with larger numbers of qubits, and the preparation of other entangled states in the $W$ and GHZ classes, to explore how these factors influence the structure and efficiency of optimal entangling protocols. The approach based on Pontryagin's minimum principle, developed here, can impact these generalizations.

\acknowledgements
This work was supported by the National Science Foundation Award No. DMR-1945395. AR is grateful to the
Kavli Institute for Theoretical Physics for hospitality during
the completion of this work, supported in part by the National
Science Foundation under Grant No. PHY-1748958.

\bibliography{references}

\end{document}


\title{Supplementary Material for ''Optimal preparation of the $W$ state for qubits with XY coupling''}

\author{Dalton Jones}
\affiliation{Department of Physics and Astronomy, University of California, Los Angeles, California 90095, USA}
\affiliation{Department of Physics and Astronomy, Western Washington University, Bellingham, Washington 98225, USA}

\author{Armin Rahmani}
\affiliation{Department of Physics and Astronomy, Western Washington University, Bellingham, Washington 98225, USA}
\affiliation{Advanced Materials Science and Engineering Center, Western Washington University, Bellingham, Washington 98225, USA}
\affiliation{Kavli Institute for Theoretical Physics, University of California, Santa Barbara, California
93106, USA}

\date{\today}

\begin{abstract}

This supplement provides a review of Pontryagin's formalism, details on the singular interval, and a discussion of dephasing and leakage.

\end{abstract}

\maketitle

\section{Review of Pontryagin's Minimum Principle}
Pontryagin's minimum principle provides remarkable insight into the mathematical structures of the protocols that minimize the cost function for system evolving with a differential equation containing the controls. We will use it as a means of verifying and improving our numerical results. To employ Pontryagin's minimum principle for a quantum system evolving with the Schrodinger equation, we first construct a mathematical object known as the optimal-control Hamiltonian (OCH), not to be confused with the quantum Hamiltonian that generates the physical time evolution. The OCH is defined as 
\begin{eqnarray}\label{eq:MCH}
{\cal H} &=& {\bf p}^{\rm T}.{\bf f(\bf x,\{\alpha\}} ),  
\end{eqnarray}
for a system with dynamical variables $\bf x$ evolving with the equations of motion
\begin{equation}\label{eq:EOM0}
\dot{\bf x}={\bf f(\bf x,\{\alpha\}} ),
\end{equation}
where $\{\alpha(t)\}$ is the set of control functions, and $\bf p$ are conjugate momenta defined with respect to the dynamical variables $\bf x$. We treat $\bf x$ and $\bf p$ as column vector, with the superscript $\small{\rm T}$ indicating the matrix transpose. The conjugate momenta evolve with the Hamilton equations of motion
\begin{equation}\label{eq:EOM}
\dot{\bf p}=-\partial_{\bf x}{ \cal H}.
 \end{equation}
The analogous $\dot{\bf x}=\partial_{\bf p}{ \cal H}$ is equivalent to Eq.~\eqref{eq:EOM0}.
The conjugate momenta satisfy the following boundary condition:
\begin{equation}\label{eq:pBC}
{\bf p(\tau)}=\partial_{\bf x}{\cal E({\bf x(\tau)})},
\end{equation}
where $\cal E$ is the cost function to be minimized.

If we insert the optimal trajectories ${\bf x}_{\rm opt}$ and ${\bf p}_{\rm opt}$ corresponding to the optimal protocol into the OCH, it becomes a function of the control parameters $\{ \alpha\}$. The optimal $\alpha_{\rm opt}$ then minimizes the OCH at every point in time
\begin{equation}
{\cal H}({\bf x}_{\rm opt},{\bf p}_{\rm opt},\{ \alpha_{\rm opt}\})=\min_{\{ \alpha\}}{\cal H}({\bf x}_{\rm opt},{\bf p}_{\rm opt},\{ \alpha\}).
 \end{equation}

The state of a quantum system can be represented by a complex vector $|\psi(t)\rangle={\bf R}+i{\bf I}$, where ${\bf R}$ and ${\bf I}$ are the real and imaginary parts of the quantum state. For a real Hamiltonian, the Schr\"odinger equation reduces to the following equations of motion for the dynamical variables $\bf R$ and $\bf I$:
\begin{equation}\label{eq:eff}
\dot{\bf R}=H(t){\bf I}, \quad  \dot{\bf I}=-H(t){\bf R}. 
\end{equation}

The OCH can then be written as
\begin{eqnarray}\label{eq:explicitH}
{ \cal H} &=& {\bf P_R}^{\rm T} H(t) {\bf I}-{\bf P_I}^{\rm T} H(t) {\bf R},
\end{eqnarray}
where ${\bf P}_{\bf R}$  and ${\bf P}_{\bf I}$ respectively denote the conjugate momenta to the dynamical variable vectors $\bf R$ and $\bf I$. 
By combining the above real conjugate momenta into a complex vector
\begin{equation}
|\Pi\rangle={\bf P_R}+i{\bf P_I},
\end{equation}
we can write the OCH as \begin{equation}{ \cal H}= {\rm Im}[\langle\Pi(t)| H(t) |\psi(t)\rangle].\label{eq:OCH2}
\end{equation}

\section{Pontryagin coefficient in the singular interval and convergence time}\label{app:a}

The optimal protocols found in the main text have a structure, where only a field $B_3$ is turned on Thus, the state can be written as 
\begin{eqnarray}
|\psi(t)\rangle&=&U_1(t)|\psi_0\rangle, \quad t<t_1,\\
|\psi(t)\rangle&=&U_2(t-t_1)U_1(t_1)|\psi_0\rangle,\quad t_1<t<t_2,\nonumber\\
|\psi(t)\rangle&=&U_1(t-t_2)U_2(t_2-t_1)U_1(t_1)|\psi_0\rangle,\quad t_2<t<\tau,\nonumber
\end{eqnarray}
where $U_j(t) = e^{-itH_j}$ for $j=1,2$ and
\begin{eqnarray}
H_1 &\equiv& H(B_1=B_2=J_{12}=0, B_3=1, J_{23}=J_{31}=1), \\
H_2 &\equiv& H(B_1=B_2=J_{12}=0, B_3=1, J_{23}=J_{31}=0).
\end{eqnarray}
We then find that $U_2$ is a diagonal matrix $U_2(t)={\rm diag}(e^{-it},e^{it},e^{it})$ and $U_1(t)$ is given by the matrix
\begin{equation*}
\left(\begin{array}{ccc}
c_3+{1\over 3}is_3 & {2\over 3}is_3 & {2\over 3}is_3  \\ 
{2\over 3}is_3 & {1\over 2}c_1+{1\over 2}c_3-{1\over 2}is_1-{1\over 6}is_3  &  -2c_1c_1^2+{2\over 3}ic_1^3 \\ 
{2\over 3}ic_3 & -2c_1s_1^2+{2\over 3}is_1^3 &  {1\over 2}c_1+{1\over 2}c_3-{1\over 2}is_1-{1\over 6}is_3
\end{array} \right),
\end{equation*}
where we have introduced the shorthand notation $c_n\equiv\cos (nt)$ and $s_n\equiv\sin (nt)$.
We can then similarly write the conjugate momenta as
\begin{eqnarray}
|\Pi(t)\rangle&=&-{2\over3} U_1^\dagger(\tau-t){\cal M}|\psi(\tau)\rangle, \quad t_2<t<\tau,\\
|\Pi(t)\rangle&=&-{2\over3} U_2^\dagger(t_2-t)U_1^\dagger(\tau-t_2){\cal M}|\psi(\tau)\rangle, \quad t_1<t<t_2,\nonumber\\
|\Pi(t)\rangle&=&-{2\over3} U_1^\dagger(t_1-t)U_2^\dagger(t_2-t_1)U_1^\dagger(\tau-t_2){\cal M}|\psi(\tau)\rangle,\quad t<t_1.\nonumber
\end{eqnarray}
The above expressions for the quantum state and the conjugate momenta allow us to calculate $\partial_J{\cal H}$ and $\partial_B{\cal H}$ as a function of time, using Eq.~\eqref{eq:OCH2} and its analogs for other control parameters.

Focusing on the singular interval $t_1<t<t_2$, let us, as an example, determine $\partial_{J_{23}}{\cal H}={\rm Im}[\langle\Pi(t)| {\cal K}_{J_{23}} |\psi(t)\rangle]$, where ${\cal K}_{\alpha} =\partial_\alpha H$ leading to
\begin{eqnarray*}
\langle\Pi(t)| {\cal K}_{J_{23}} |\psi(t)\rangle=
-{2 \over 3} \langle\psi(0)|U_1^\dagger(t_1)U_2^\dagger(t_2-t_1)U_1^\dagger(\tau-t_2) \\
{\cal M}U_1(\tau-t_2)U_2(t_2-t){\cal K}_{J_{23}}U_2(t-t_1)U_1(t_1)|\psi(0)\rangle
\end{eqnarray*}
We can explicitly write the above expression by using 
\begin{equation}
{\cal K}_{J_{23}} =\left(\begin{array}{ccc}
0&-2&0 \\ 
-2 &0 & 0 \\ 
0 & 0&0
\end{array} \right),\quad  |\psi(0)\rangle=\left(\begin{array}{c}
1 \\ 
0 \\ 
0
\end{array} \right).
\end{equation}
The imaginary part of $\langle\Pi(t)| {\cal K}_{J_{23}} |\psi(t)\rangle$ above is a lengthy trigonometric expression, which can be calculated explicitly. It turns out that when using the values of $t_1$ and $\tau_0$ from the bang-bang optimization, this function vanishes for all $t$ and $\tau$, indicating the singularity of the interval.
It is important to emphasize that the above analytical expressions depend on the particular bang-bang form of the optimal protocol, which we obtained through numerical optimization. However, we were able to use these analytical expressions to fine tune the values of $t_1$ and $\tau_0$ with extreme precision. Using these refined values, we wrote the error $\cal E$ in terms of $\tau$ using the overlap
\begin{equation}
\langle \psi_{\rm target}|\psi(\tau)\rangle=\langle \psi_{\rm target}|U_1(\tau_0-t_1)U_2(\tau-\tau_0)U_1(t_1)|\psi_0\rangle,
\end{equation}
then, after setting the error to zero, we found the smallest value of $\tau=\tau_C = 0.7727$ at which the resulting lengthy trigonometric expression vanishes.

\section{Relaxation, dephasing, and leakage}

As an example, we investigate the implementation of the protocols within the superconducting gmon qubit architecture. Typical values of relaxation and decoherence times for gmon qubits are on the order of $10 \ \mu s = 10^{-5}\ s \ $~\cite{Mart,gmonrelax}. The time scale of our problem $\tau_C$, which is of order $1$ in natural units, can be compared to these times by converting back to SI units. This is done using the typical energy scale for a gmon qubit, $\Lambda \approx 2\pi\ GHz \ $~\cite{Mart,gmonrelax,Spect}, such that the time scale of our protocols is $\tau_{SI} = \tau_C/\Lambda \approx 10^{-10}\ s \ $. Therefore, we see that our protocols operate on time scales that are approximately five orders of magnitude smaller than typical relaxation and dephasing times, such that their contributions to the final error will be negligible.

One important concern is leakage out of the qubit subspace, particularly due to sharp transitions, which excite high-frequency modes. The robustness of the optimal bang-bang protocols to timing errors suggests that we can eliminate high frequencies by constructing smooth ramps in place of the sharp jumps over the timescales where errors remain negligible.

Here, we consider a particular model of leakage in some detail, which still suggests bang-bang protocols, but requires longer time scales. However, we believe that this minimal model would also break in the presence of sharp transitions, and replacing the jumps with smooth ramps over small timescales is a better practical approach.

\begin{figure}
    \centering
    \includegraphics[width=0.9\linewidth]{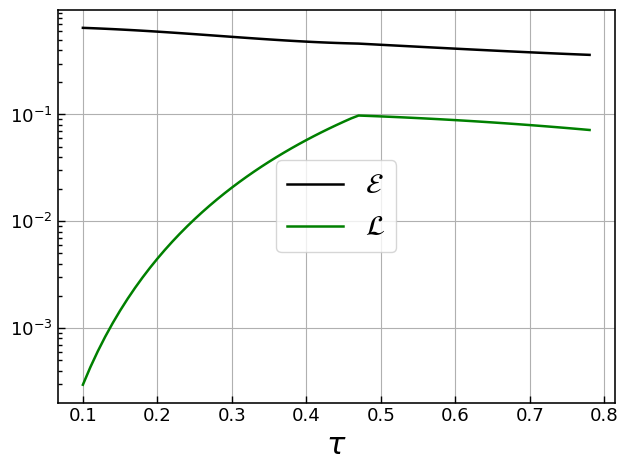}
    \caption{The error $\mathcal{E}$ between the final state of the system and the desired $W$ state and leakage $\mathcal{L}$ outside of the qubit basis for the bang-bang protocols at various final times ($0\leq \tau \leq 0.77$).}
    \label{fig:leakage}
\end{figure}

\begin{figure*}
    \centering
    \includegraphics[width=0.9\linewidth]{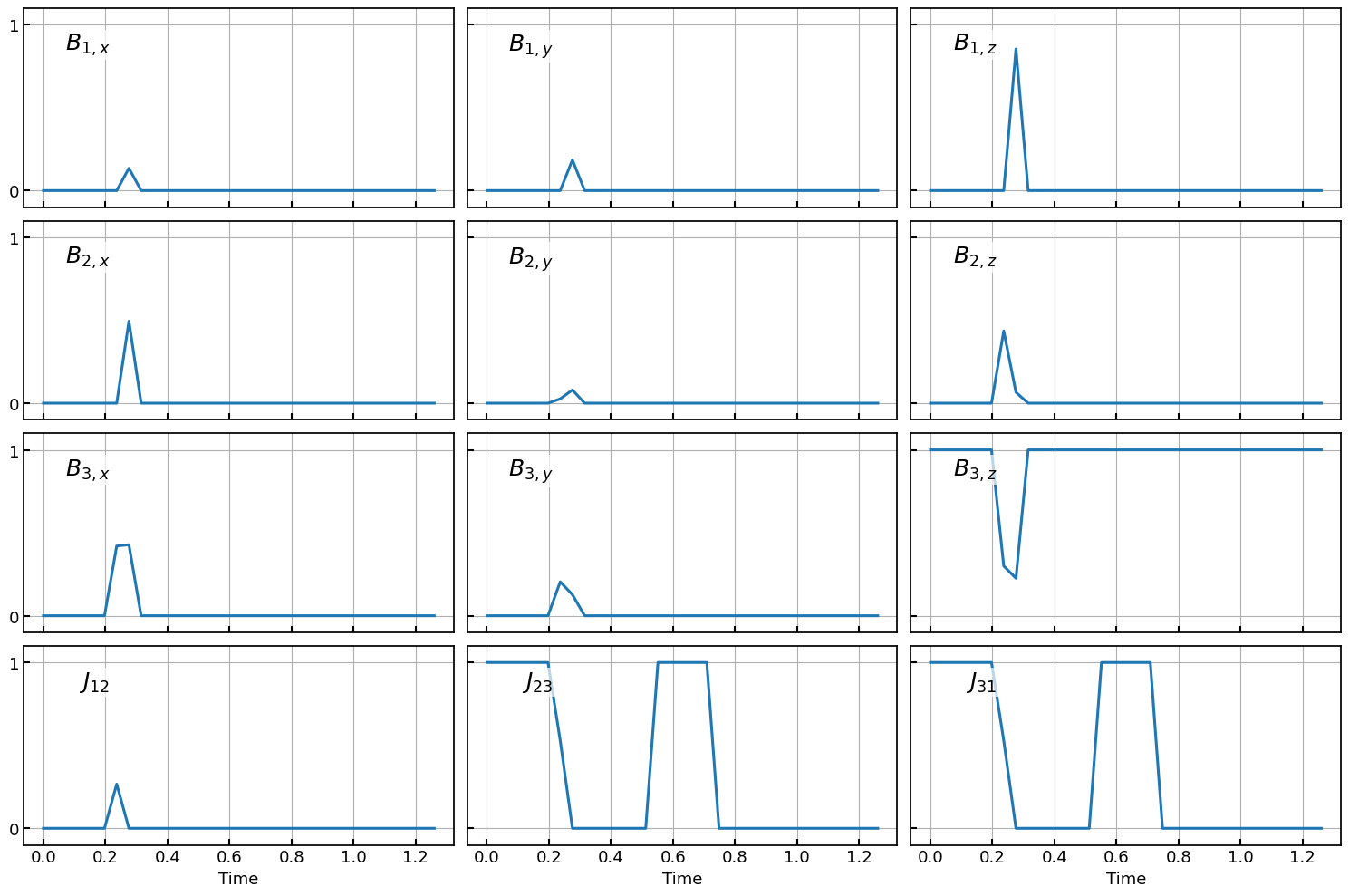}
    \caption{Optimal protocols for the $B$ and $J$ control parameters with final time $\tau=1.3$, determined by an $N$-interval optimization ($N=33$) of the cost function $\mathcal{C}$, leading to $\mathcal{E} \approx 0.276$ and leakage $\mathcal{L} \approx 0.0726$.}
    \label{fig:Optleak}
\end{figure*}

Errors resulting from leakage outside the qubit basis during bang-bang protocols are expected to be significant, especially due to the weak anharmonicities of gmon qubits~\cite{gmonrelax, Mart2}. We model leakage to energy levels $|n\rangle$, where $n\geq2$, by first generalizing our Hamiltonian according to Refs.~\cite{Spect,Mart,Duff} which map the Pauli matrices to ladder operators $\sigma^x \rightarrow (a+a^\dagger)$, $\sigma^y \rightarrow i(a-a^\dagger)$, and $\sigma^z \rightarrow a^\dagger a$. The Hamiltonian takes the form of three coupled Duffing oscillators written in terms of the single qubit ladder operators $a_i$ ($a_i^\dagger$) with anharmonic terms $- \frac{\alpha_i}{2}a_i^\dagger a_i^\dagger a_ia_i$ for each qubit.
\begin{eqnarray}
    H_i  &=& 2B_{i,z} a_i^\dagger a_i - \frac{\alpha_i}{2}a_i^\dagger a_i^\dagger a_ia_i + B_{i,x}(a_i+a_i^\dagger) - i B_{i,y}(a_i-a_i^\dagger),  \nonumber \\
    H_{m,n} &=& J_{mn}(a_m^\dagger a_n + a_m a_n^\dagger),  \nonumber \\
    H &=& \sum_i^3 H_i - \sum_{\langle m,n\rangle}H_{m,n}.
\end{eqnarray}
Since our goal is to model the leakage for our optimal protocols, our Hamiltonian is simplified by setting the $x$ and $y$ fields to zero as well as $J_{12}, B_{1,z}, \text{ and } B_{2,z}$
\begin{eqnarray}
    H  &=& 2B_{3} a_3^\dagger a_3 - \frac{\alpha}{2}a_3^\dagger a_3^\dagger a_3a_3   \nonumber \\
    && -J_{23}(a_2^\dagger a_3 + a_2 a_3^\dagger) - J_{31}(a_1^\dagger a_3 + a_1 a_3^\dagger).\label{eq:leakageH}
\end{eqnarray}
Considering that the state of our system evolves within the $\sum_i m_i = +1$ qubit basis sector, the leakage transitions during evolution arise from the coupling terms
\begin{eqnarray}
    a_1 a_3^\dagger |101\rangle &= \sqrt{2} |002\rangle,  \ \ \quad \ \ 
    a_1^\dagger a_3 |101\rangle =& \sqrt{2} |200\rangle, \nonumber \\
    a_2 a_3^\dagger |011\rangle &= \sqrt{2} |002\rangle, \ \  \quad \ \ 
    a_2^\dagger a_3 |011\rangle =& \sqrt{2} |020\rangle,  \label{leakage}
\end{eqnarray}
where $|0\rangle = |\downarrow\rangle$ and $|1\rangle = |\uparrow\rangle$. These transitions and their Hermitian conjugates (e.g. $a_1^\dagger a_3|002\rangle = \sqrt{2} |101\rangle$) contribute to leakage outside of the qubit subspace and increase the error in preparing a state in the qubit subspace. Limiting the leakage model to the leading case where at most one qubit is in the $n=2$ state, we define leakage as 
\begin{equation}\label{eq:leakage}
    \mathcal{L} = P_{0,0,2}(\tau) + P_{0,2,0}(\tau) + P_{2,0,0}(\tau) ,
\end{equation}
where, for example, $P_{0,0,2}(\tau) = |\langle \psi(\tau)|0,0,2\rangle|^2$. The Hamiltonian basis may also be truncated to the first three single-qubit energy levels $|0\rangle, |1\rangle, |2\rangle$ without loss of generality
\begin{eqnarray}
H &=& 2B_{3}|1\rangle \langle 1| + (4B_{3} - \alpha_3) |2\rangle \langle 2| - H_{2,3} - H_{3,1},  \\ \nonumber
H_{m,n} &=& J_{mn} \Big[\Big(|1\rangle \langle 0| + \sqrt{2}|2\rangle \langle 1| \Big) \otimes \Big(|0\rangle \langle 1| + \sqrt{2}|1\rangle \langle 2| \Big) +  {\rm H.c.} \Big].
\end{eqnarray}

The parameters are written in units normalized by the energy scale of the system such that a reasonable value of the anharmonicity in these units is $\alpha_3 \approx 300\ MHz/ 1\ GHz = 0.3$~\cite{gmonrelax, Mart2}. As expected, evolving the system according to the bang-bang protocols found in the ideal case leads to a significant increase in error and leakage shown in Fig.~\ref{fig:leakage}. Interestingly, the time $\tau_0 = 0.469017$ at which $J_{23}(t)=J_{31}(t)$ transitions from a protocol without switches ($t_2=t_1$) to a protocol with two switches ($t_2>t_1$) also corresponds to the time at which leakage behavior abruptly transitions from increasing to decreasing. This signifies the potential to minimize leakage and error while retaining the bang-bang structure. In fact, the Hamiltonian that includes leakage Eq.~\ref{eq:leakageH} is still linear in the control parameters, such that Pontryagin's minimum principle dictates that protocols which optimize error and leakage will generically be bang-bang. However, optimal solutions may have a hybrid pulse-shape optimization and bang-bang structure if there are intervals over which the coefficients of control parameters in the OCH are singular.

Here, we perform a numerical optimization of the weighted sum of the error and leakage \eqref{eq:leakage}
\begin{equation}\label{eq:totalcost}
    \mathcal{C}(\tau) = \mathcal{E}(\tau) + w_\mathcal{L} \mathcal{L}(\tau),
\end{equation}
where we choose a weight of $w_\mathcal{L}=0.5$ because our ultimate goal is to reach the $W$ state rather than simply minimize leakage. We performed a similar $N$-step optimization with simulated annealing using a piecewise constant protocol. For times within the range ($0\leq t \leq 0.77$) of our original protocols, we do not see solution convergence, nor any improvement in error or leakage. However, when performing the optimization for longer time scales, we begin to see solution convergence and a slight improvement in error and leakage. An example of one of these optimizations for $\tau = 1.3$ and $N=33$ is provided in Fig.~\ref{fig:Optleak}. This protocol seems to retain most of the bang-bang characteristic of the original solution while adding an additional pulse to each of the control parameter protocols. This solution error $\mathcal{E}(\tau=1.3)\approx 0.276$ and leakage $\mathcal{L}(\tau=1.3) \approx 0.0783$. Comparing this to the bang-bang protocol at $\tau = 0.77$, which has error $\mathcal{E}_{bang}(\tau=0.77) \approx 0.364$ and leakage $\mathcal{L}_{bang}(\tau=0.77) \approx 0.0726$, shows a slight improvement in error and leakage. We expect both error and leakage to continue decreasing as we allow for larger and larger final times.

Further evidence for an increase in the convergence time is that other leakage minimization methods, such as Derivative Removal by Adiabatic Gate (DRAG) and other pulse shaping methods, are limited to time scales at which natural suppression of leakage begins to occur $t_{gate} \approx 1/\alpha \approx 3.3$ due to the detuning between the frequency spectrum of the pulse and the anharmonicity~\cite{Mart2,1Qleak}, which is $4$x the time scale of our original solutions $\tau_C\approx 0.7727$.

\bibliography{references}